\documentclass[a4paper,12pt]{article}

\usepackage{amsmath,amsthm,amssymb,bbm}

\title{Squeezing Inequalities and Entanglement for Identical Particles}

\author{F. Benatti$^{a,b}$, R. Floreanini$^b$, U. Marzolino$^{a,b}$\\
\\
\small ${}^a$Dipartimento di Fisica, Universit\`a di Trieste, 
34151 Trieste, Italy\\
\small ${}^b$Istituto Nazionale di Fisica Nucleare, Sezione di Trieste,
34151 Trieste, Italy}

\date{\null}

\begin{document}

\maketitle

\begin{abstract}
\noindent
By identifying non-local effects in systems of identical Bosonic qubits through correlations of
their commuting observables, we show that entanglement is not necessary to violate certain squeezing
inequalities that hold for distinguishable qubits and that spin squeezing may not be necessary to achieve
sub-shot noise accuracies in ultra-cold atom interferometry.
\end{abstract}

\section{Introduction}

A generic state of $N$ distinguishable qubits is defined to be entangled
if it is not fully separable~\cite{Hororev}, namely if it cannot be written as
\begin{equation}
\label{sep,dist}
\rho_{sep}=\sum_k p_k \rho_k^{(1)}\otimes\rho_k^{(2)}\cdots\otimes\rho_k^{(N)}\ ,
\end{equation}
where the $p_k\geq 0$ are weights, $\sum_kp_k=1$, while each $\rho_k^{(j)}$ is a density matrix
for the $j$-th qubit acting on the corresponding Hilbert space $\mathbbm{H}_j$.

Entangled $N$-qubit states have been proposed as means to beat the so-called shot-noise limit
in metrological applications~\cite{Kitagawa1993,Wineland1994} and spin-squeezing techniques
have been devised to generate them in systems of ultra-cold atoms~\cite{Oberthaler1,Treutlein}.
These states are then used in interferometric experiments where their states are rotated
by means of collective spin components; in the case of distinguishable qubits,
all these rotations are local; therefore,  preliminary spin squeezing of separable states is necessary
in order to introduce (non-local) quantum correlations.

However, in the case of ultra-cold trapped atoms, the qubits involved are identical, a fact that
asks for a rethinking of the properties and behaviors valid for distinguishable qubits.
Indeed, the notion of separability based on~(\ref{sep,dist}) is strictly associated with the tensor
product structure of the Hilbert space, $\mathbbm{H}_N=\bigotimes_{j=1}^N\mathbbm{H}_j$, which is natural
for $N$ distinguishable particles. On the other hand, pure Bosonic states must be symmetric under exchange
of particles and mixed states must be convex combinations
of projections onto such states. This fact demands a different approach to the notions of non-locality
and entanglement based not on a structure related to the particle aspect of first quantization,
as in~(\ref{sep,dist}), rather on the behavior of correlation functions of commuting
observables~\cite{Za,Na,Viola,Benatti2010},
more generally related to the mode description  typical of second quantization.

In the following, based on a generalized notion of entanglement~\cite{Benatti2010} which reduces to the
standard one for distinguishable qubits, we show that neither entangled states, nor spin-squeezing are
necessary in order to achieve sub-shot noise accuracies. Though necessary, non-locality comes not from the
states, rather from the rotations that are implemented in the interferometric experiments.
Indeed, while all rotations are local for distinguishable qubits, certain collective spin observables become
instead non-local for identical qubits and might permit sub-shot noise accuracies without the need
of a preliminary squeezing of the input state as in~\cite{Oberthaler1,Treutlein}.

\section{Entanglement for Identical Particles}

In general, density matrices as in~\eqref{sep,dist} are not allowed Bosonic states not even if
$\rho^{(i)}_k=\rho_k$ for all $i$.
Indeed, consider two qubits and fix an orthonormal basis $\mathbb{C}^4\ni\vert ij\rangle$,
$i,j=\uparrow,\downarrow$. If they are Bosons, their states cannot have non-vanishing components on the
anti-symmetric state
$\displaystyle\vert\Psi_-\rangle=\frac{\vert\downarrow\uparrow\rangle-\vert\uparrow\downarrow\rangle}
{\sqrt{2}}$. Since
$\displaystyle
\langle\Psi_-\vert\rho\otimes\rho\vert\Psi_-\rangle=\hbox{Det}(\rho)$,
the density matrix $\rho\otimes\rho$ cannot correspond to a state of two Bosonic qubits
unless $\rho$ is a projection ($\hbox{Det}(\rho)=0$).

Therefore, the tensor product structure which is natural for distinguishable particles is not appropriate
for discussing the entanglement properties of systems of indistinguishable particles. These should rather
be investigated within the second quantization formalism whereby one introduces creation and annihilation
operators $a^\#_i$ of single-particle orthonormal basis states $\vert i\rangle$, obeying the canonical
commutation relations $[a_i\,,\,a^\dag_j]=\delta_{ij}$.

Entanglement in such a context should correspond to whether, given a state $\omega$ of the system, there
are non-classical correlations among commuting observables~\cite{Na,Benatti2010}; for instance,
between non-overlapping spatial regions $V_{1,2}$.
Specifically, this can be inspected by considering the structure of two-point functions of the form
$\omega(P_{V_1}P_{V_2})$, where $P_{V_\alpha}$ is any polynomial in creation and annihilation operators
$a^\#(\psi_\alpha)$ of states $\vert\psi_\alpha\rangle$ spatially localized in the region $V_\alpha$.
The relevant fact is that the spatially local algebras generated by polynomials $P_{V_\alpha}$ commute:
$[a(\psi_1)\,,\,a^\dag(\psi_2)]=\langle\psi_1\vert\psi_2\rangle=0$.

More in general, one can argue about the entanglement between observables belonging to two generic commuting
sub-algebras $(\mathcal{A},\mathcal{B})$ of the entire algebra generated by creation and annihilation operators,
which we shall refer to as \textit{algebraic bipartition}~\cite{Benatti2010}.
We shall call an operator $(\mathcal{A},\mathcal{B})$-local if of the form $AB$, $A\in\mathcal{A}$ and
$B\in\mathcal{B}$ and a state $\omega$ $(\mathcal{A},\mathcal{B})$-separable
if the expectations $\omega(AB)$ of local operators $AB$ can be decomposed into a convex linear combination
of product of expectations:
\begin{equation}
\label{def-sep}
\omega(AB)=
\sum_i\lambda_i\omega^a_i(A)\,\omega^b_i(B)\ ,\quad
\lambda_i>0\ ,\ \sum_i\lambda_i=1\ ,
\end{equation}
in terms of other states $\omega^{a,b}_i$; otherwise, $\omega$ is $(\mathcal{A},\mathcal{B})$-entangled.
\medskip

\noindent
\textbf{Remark 1.}\quad
In the case of two qubits, the above definitions reproduce the standard notions
if one chooses the algebraic bipartition $\mathcal{A}=\mathcal{B}=M_2$, where $M_2$ the algebra
of $2\times 2$ matrices over $\mathbbm{C}^2$, and
$\omega(AB)={\rm Tr}\left(\rho\,A\otimes B\right)$, with $\rho$ a two-qubit density matrix.
However, for identical particles, there is no a priori given bi-partition so that questions about entanglement
and separability, non-locality and locality are meaningful only with reference to a
specific class of (commuting) observables.

\subsection{$N$ Bosons in a Double-Well Potential}
\label{sepAB-sec}

A concrete application of the previous considerations is the second quantization of a single-particle with
Hilbert space $\mathbbm{C}^2$ which, in the Bose-Hubbard approximation, effectively describes $N$ ultra-cold atoms
confined by a double-well potential.
Then, the state $\vert\downarrow\rangle$ describes one atom located within the left well and the state $\vert\uparrow\rangle$ an atom localized
within the right one.
Let $\vert0\rangle$ be the vacuum state and $a^\dag$, $b^\dag$ the creation operators of
a particle in the states $|\downarrow\rangle$ and $|\uparrow\rangle$, that is
$a^\dag\vert0\rangle=\vert\downarrow\rangle$,
$b^\dag\vert0\rangle=\vert\uparrow\rangle$.

When the total number $N$ is conserved, the symmetric Fock space of this two-mode system
is generated by $N+1$
orthonormal eigenvectors of the number operator $a^\dag a+b^\dag b$:
\begin{equation}
\label{NFock}
\vert k\rangle=\frac{(a^\dagger)^k (b^\dagger)^{N-k}}{\sqrt{k!(N-k)!}}\,
|0\rangle, \quad 0\leq k\leq N\ .
\end{equation}
Because of the orthogonality of the spatial modes, by considering the norm-closures of
all polynomials $P_a$ in $a,a^\dag$, respectively $P_b$ in $b,b^\dag$,
one obtains two commuting subalgebras $\mathcal{A}$ and $\mathcal{B}$.

According to~(\ref{def-sep}), the states $\vert k\rangle$  are $(\mathcal{A},\mathcal{B})$-separable;
indeed, they are created by the ($\mathcal{A},\mathcal{B}$)-local operators $(a^\dagger)^k (b^\dagger)^{N-k}$.
More in general, ($\mathcal{A},\mathcal{B}$)-separable states must be convex combinations of projections
$\vert k\rangle\langle k\vert$~\cite{Benatti2010}:
\begin{equation}
\label{sep}
\rho=\sum_{k=0}^N p_k \vert k\rangle\langle k\vert\qquad p_k>0\ ,\ \sum_{k=0}^Np_k=1\ .
\end{equation}
Consider instead the following operators
\begin{equation}
\label{jsAB}
J_x=\frac{1}{2}(a^\dag b+ab^\dag)\ ,\
J_y=\frac{1}{2i}(a^\dag b-ab^\dag)\ ,\
J_z=\frac{1}{2}(a^\dag a-b^\dag b)\ ,
\end{equation}
that satisfy the $SU(2)$ algebraic relations
$[J_x\,,\,J_y]=i\,J_z$.
They are all non-local with respect to the algebraic bipartition $(\mathcal{A},\mathcal{B})$
and such are the exponentials
$\displaystyle
{\rm e}^{i\theta J_x}$ and $\displaystyle{\rm e}^{i\theta J_y}$,
while
$\displaystyle{\rm e}^{i\theta J_z}={\rm e}^{i\theta a^\dag a}{\rm e}^{-i\theta b^\dag b}$
is $(\mathcal{A},\mathcal{B})$-local.

By means of a Bogolubov transformation to other creation and annihilation operators $(c^\#,d^\#)$,
such that $\displaystyle a=\frac{c+d}{\sqrt{2}}$ and $\displaystyle b=\frac{c-d}{\sqrt{2}}$,
one obtains another bipartition $(\mathcal{C},\mathcal{D})$ and rewrites
$$
J_x=\frac{1}{2}(c^\dag c-d^\dag d)\ ,\
J_y=\frac{1}{2i}(d^\dag c-dc^\dag)\ ,\
J_z=\frac{1}{2}(c^\dag d+ cd^\dag)\ .
$$
Relatively to $\{\mathcal{C},\mathcal{D}\}$, it is now
$\displaystyle{\rm e}^{i\theta\, J_x}={\rm e}^{i\theta\, c^\dag c}{\rm e}^{-i\theta\, d^\dag d}$
which acts locally.
\medskip

\noindent
\textbf{Remark 2.}\quad
In first quantization, an  ($\mathcal{A},\mathcal{B}$)-separable state for a $N=2$ Bosonic qubits
like $\vert 11\rangle=a^\dag b^\dag\vert 0\rangle$ corresponds to
$\displaystyle
\frac{\vert\uparrow\downarrow\rangle+\vert\downarrow\uparrow\rangle}{\sqrt{2}}$.
Such state is surely entangled for distinguishable qubits, while, according to our definition, it is no longer so
for identical Bosonic qubits; the reason is that its entanglement is only formal as it comes from the necessary
symmetrization of the separable state $\vert \uparrow\downarrow\rangle$~\cite{Ghirardi2002}.

A Bogolubov transformation as the one above corresponds to a change of basis in the single particle Hilbert
space, from the one of spatially localized states, to the one of
$\displaystyle
c^\dag\vert0\rangle=\frac{1}{\sqrt{2}}(\vert\downarrow\rangle+\vert\uparrow\rangle)$,
$\displaystyle
d^\dag\vert0\rangle=\frac{1}{\sqrt{2}}(\vert\downarrow\rangle-\vert\uparrow\rangle)$.
Physically speaking, such states are eigenstates of the single particle
Hamiltonian in the Bose-Hubbard approximation with a highly penetrable barrier.
The change to the \textit{energy bipartition} $(\mathcal{C},\mathcal{D})$
is non-local with respect to the \textit{spatial bipartition}
$(\mathcal{A},\mathcal{B})$, though it corresponds to a local unitary rotation in first quantization.

\subsection{Collective Spin Inequalities and Entanglement}

In the case of a system of $N$ distinguishable qubits, the collective angular momentum operators $J_{x,y,z}$ and
the corresponding rotations are sums of single particle spin operators, $J^{(j)}_{x,,y,z}$, {\it i. e.}
$J_{x,y,z}=\sum_{j=1}^NJ^{(j)}_{x,y,z}$. These operators are
local with respect to the tensor product structure in~\eqref{sep,dist}.

Based on this, the variance $\Delta^2 J_{\vec{n}}$ of the collective spin
$J_{\vec{n}}=n_x\,J_x+n_y\,J_y+n_z\,J_z$ along the unit spatial direction $\vec{n}=(n_x,n_y,n_z)$,
with respect to separable vector states $\vert\Psi\rangle=\bigotimes_{j=1}^N\vert\psi_j\rangle$ results
\begin{eqnarray}
\nonumber
\Delta^2 J_{\vec{n}}&=&\langle\Psi\vert J^2_{\vec{n}}\vert\Psi\rangle-\langle\Psi\vert
J_{\vec{n}}\vert\Psi\rangle^2\\
\label{spinv}
&=&\frac{N}{4}-\sum_{j=1}^N\,\Big(\langle\psi_j
\vert J^{(j)}_n\vert\psi_j\rangle\Big)^2\leq\frac{N}{4}\ .
\end{eqnarray}
Therefore, $\Delta^2 J_{\vec{n}}$ is an entanglement witness for pure states, in the sense that
if $\Delta^2 J_{\vec{n}}>N/4$ then the pure state $\vert\Psi\rangle$ cannot be fully separable.

This is no longer the case for $N$ identical Bosonic qubits.
Indeed, consider the number states $|k\rangle$ in~(\ref{NFock});
using~(\ref{jsAB}), one gets
\begin{eqnarray}
\label{FI1}
\langle k\vert J_{\vec{n}}\vert k\rangle&=&\frac{n_z}{2}\,(2k-N)\\
\nonumber
\langle k\vert J^2_{\vec{n}}\vert k\rangle&=&\frac{N+2k(N-k)}{4}\\
\label{FI2}
&+&
n_z^2\frac{N(N-1)-6k(N-k)}{4}\\
\label{FI3}
\Delta^2\, J_{\vec{n}}&=&\frac{1-n^2_z}{4}\,\Big(N+2k(N-k)\Big)\ .
\end{eqnarray}
Therefore, if $k\neq 0,N$, for all $\vec{n}$ that satisfy 
$$
n^2_z<\frac{2k(N-k)}{N+2k(N-k)}\leq 1\ ,
$$
the states $\vert k\rangle$, though
$(\mathcal{A},\mathcal{B})$-separable, nevertheless yield $\Delta^2\, J_{\vec{n}}>N/4$;
therefore, $\Delta^2\, J_{\vec{n}}$ is not an entanglement witness for pure states of Bosonic qubits.

In greater generality, inequalities for mean values and variances of collective spin operators
with respect to any (mixed) separable state of distinguishable
qubits~(\ref{sep,dist}) have been derived in~\cite{Toth2009};
these are called \textit{spin squeezing inequalities} and read~\footnote{In~\cite{Toth2009}, these inequalities
are derived with respect to the standard triplet $\vec{n_1}=\hat{x}$, $\vec{n}_2=\hat{y}$,
$\vec{n}_3=\hat{z}$. The result easily extends to more general triplets.}
\begin{eqnarray}
\hskip -1.5cm
&& 
\langle J_{\vec{n}_1}^2\rangle+\langle J_{\vec{n}_2}^2\rangle+\langle J_{\vec{n}_3}^2
\rangle-\frac{N(N+2)}{4}\leq 0\ , \label{spinineq1} \\
\hskip -1.5cm
&& 
\Delta^2 J_{\vec{n}_1}+\Delta^2 J_{\vec{n}_2}+\Delta^2 J_{\vec{n}_3}-\frac{N}{2}\geqslant 0\ ,
\label{spinineq2} \\
\hskip -1.5cm
&& 
\langle J_{\vec{n}_1}^2\rangle+\langle J_{\vec{n}_2}^2\rangle-\frac{N}{2}-(N-1)\Delta^2
J_{\vec{n}_3}\leq 0\ , \label{spinineq3} \\
\hskip -1.5cm
&& 
(N-1)(\Delta^2 J_{\vec{n}_1}+\Delta^2 J_{\vec{n}_2})-\langle J_{\vec{n}_3}^2\rangle-\frac{N(N-2)}{4}\geqslant 0\ , \label{spinineq4}
\end{eqnarray}
where $\vec{n}_{1,2,3}$ denotes any triplet of unit vectors corresponding to orthogonal
spatial directions and $\langle X\rangle$ the mean value of an operator $X$.

It is thus interesting to study whether these inequalities are also satisfied by
$(\mathcal{A},\mathcal{B})$-separable states~(\ref{sep}) of $N$ identical qubits.
Let $\langle k^a\rangle=\sum_{k=0}^Np_k\,k^a$, $a=1,2$, denote
first and second moments of the $N+1$-valued stochastic
variable $k$ with respect to the probability distribution $\pi=\{p_k\}_{k=0}^N$.
Using~(\ref{FI1}),~(\ref{FI3}), mean-values and variances of collective spin
operators $J_{\vec{n}}$ with respect to the states in~(\ref{sep}) read
\begin{eqnarray}
\label{varJnc}
\langle J_{\vec{n}}\rangle&=&\frac{n_z}{2}\Big(2\langle k\rangle-N\Big)\\
\nonumber
\langle J^2_{\vec{n}}\rangle&=&\frac{N(1+2\langle k\rangle)-2\langle k^2\rangle}{4}\\
\label{varJnd}
&+&\frac{n^2_z}{4}
\Big(N(N-1)-6N\langle k\rangle+6\langle k^2\rangle\Big)\\
\nonumber
\Delta^2J_{\vec{n}}&=&\frac{N(1+2\langle k\rangle)-2\langle k^2\rangle}{4}\\
\label{varJne}
&+&\frac{n^2_z}{4}
\Big(6\langle k^2\rangle-2\langle k\rangle(N+2\langle k\rangle)-N\Big)\ .
\end{eqnarray}
From the orthogonality of the triplet $\vec{n}_{1,2,3}$, it follows that $n_{1z}^2+n_{2z}^2+n_{3z}^2=1$;
one can thus check that all inequalities but~(\ref{spinineq3}) are satisfied by
$(\mathcal{A},\mathcal{B})$-separable states. Concerning~(\ref{spinineq3}), its left hand side reads
\begin{eqnarray}
\nonumber
\delta&=&\frac{N}{2}\Big(\Delta^2k-\langle k\rangle(N-\langle k\rangle)\Big)\\
\label{idsqueez1}
&+&
\frac{n_{3z}^2}{2}\Big((N+2)\langle k\rangle(N-\langle k\rangle)-3N\Delta^2_k\Big)\ ,
\end{eqnarray}
where $\Delta^2k:=\langle k^2\rangle-\langle k\rangle^2$ is the
variance of $k$ with respect to $\pi=\{p_k\}_{k=0}^N$.
If $\pi$ is chosen such that $a:=\langle k\rangle(N-\langle k\rangle)>N\Delta^2k$,
then $\delta$ becomes positive and thus~(\ref{spinineq3}) results violated by the corresponding
$(\mathcal{A},\mathcal{B})$-separable states~(\ref{sep}) for all orthogonal triplets with
\begin{equation}
\label{idsqueez2}
1\geq n^2_{3z}\,>\,\frac{N(a-\Delta^2k)}{(N+2)\,a-3\,N\,\Delta^2k}\ .
\end{equation}
Consider the pure states $\vert \ell\rangle$ in~(\ref{NFock}) with
$\ell\neq 0,N$; in such a case, $p_k=\delta_{k\ell}$, $\langle k\rangle=\ell$ and $\Delta^2 k=0$, so
that~(\ref{spinineq3}) is violated for $1\geq n^2_{3z}>N/(N+2)$.

\section{Spin Squeezing and Metrology}

The preceding results indicate that spin-squeezing inequalities that are derived
for distinguishable qubits can not directly be used as entanglement witnesses
in the context of identical qubits. Since the use of spin-squeezed states for
metrological purposes have recently become the focus of much
theoretical~\cite{Kitagawa1993,Wineland1994,Soerensen2001.1,Pezze2009} investigations,
we now discuss the impact of particle indistinguishability on such an issue.

\subsection{Spin Squeezing}

For any orthogonal triplet of space-directions $\vec{n}_{1,2,3}$,
the Heisenberg uncertainty relations for the SU(2) operators $J_{\vec{n}}$ read
\begin{equation}
\Delta^2 J_{\vec{n}_1}\Delta^2 J_{\vec{n}_2}\geqslant\frac{1}{4}\langle J_{\vec{n}_3}\rangle^2 \ .
\end{equation}
One speaks of spin-squeezing when one of the variances can be made smaller than
$\frac{1}{2}\,\Big|\langle J_{\vec{n}_3}\rangle\Big|$.
The relevance of states satisfying this condition for achieving otherwise unavailable accuracies
has been studied in relation to the measure of an angle $\theta$ by interferometric techniques. These are based
on a rotation of an input state $\rho$ into
\begin{equation}
\label{state-rot}
\rho_\theta=\exp(-i\theta J_{\vec{n}_1})\,\rho\,\exp(i\theta J_{\vec{n}_1})\ ,
\end{equation}
and upon measuring on $\rho_\theta$ the collective spin $J_{\vec{n}_2}$,
where $\vec{n}_2\perp\vec{n}_1$. By choosing
the remaining orthogonal unit vector $\vec{n}_3$ such that
$\langle J_{\vec{n}_3}\rangle={\rm Tr}(\rho\, J_{\vec{n}_3})\neq 0$,
by error propagation, the uncertainty $\delta\theta$ in the determination of $\theta$ can
be estimated by~\cite{Wineland1994}
\begin{equation}
\label{Squeezparam0}
\delta^2\theta=\frac{\Delta^2 J_{\vec{n}_1}}{\Big(\partial_\theta\langle
J_{\vec{n}_2}\rangle_\theta\Big\vert_{\theta=0}\Big)^2}
=\frac{\Delta^2 J_{\vec{n}_1}}{\langle J_{\vec{n}_3}\rangle^2}=\frac{\xi_W^2}{N}\ ,
\end{equation}
in terms of the spin-squeezing parameter
\begin{equation}
\label{Squeezparam}
\xi_W^2:=\frac{N\Delta^2 J_{\vec{n}_1}}{\langle J_{\vec{n}_3}\rangle^2}\ .
\end{equation}
The value $\displaystyle\delta^2\theta=\frac{1}{N}$ is called shot-noise limit; in the case of
distinguishable qubits, it gives the lower bound to the attainable accuracies when the input
state $\rho$ is separable. Indeed, in such a case one finds $\xi^2_W\leq 1$.
This result follows from the inequality $\xi^2_W\geq \xi_2^S$ where the new spin-squeezing parameter
\begin{equation}
\label{xiS}
\xi_S^2:=\frac{N\Delta^2 J_{\vec{n}_1}}{\langle J_{\vec{n}_2}\rangle^2+\langle J_{\vec{n}_3}\rangle^2}\ ,
\end{equation}
has been introduced in~\cite{Soerensen2001.1}; by means of the local structure of the collective spin operators
$J_{x,y,z}$, one can prove that $\xi_S^2$ is always $\geq 1$ for separable states of
distinguishable qubits.
Therefore, using distinguishable qubits, the shot-noise limit can be beaten, namely
accuracies better than $1/N$ can be achieved only if $\xi^2_S<1$, that is only by means of entangled states.

Let us instead consider $N$ identical Bosonic  qubits in the $(\mathcal{A},\mathcal{B})$-separable pure
states $\vert k\rangle$ and any triplet of orthogonal spatial directions $\vec{n}_{1,2,3}$ with
$\vec{n}_1\neq \hat{z}$.
Using~(\ref{FI1})-(\ref{FI3}) and $n^2_{1z}+n^2_{2z}+n^2_{3z}=1$, one computes
\begin{eqnarray}
\nonumber
\xi_S^2&=&\frac{N\Delta^2 J_{\vec{n}_1}}{\langle J_{\vec{n}_2}\rangle^2
+\langle J_{\vec{n}_3}\rangle^2}= N\,
\frac{1-n^2_{1z}}{n^2_{2z}+n_{3z}^2}\,\frac{N+2k(N-k)}{(2k-N)^2}\\
&=&
\frac{N(N+2k(N-k))}{(2k-N)^2}\geq 1\quad, \quad 0\leq k\leq N\ .
\label{xiS1}
\end{eqnarray}
In the case of the $(\mathcal{A},\mathcal{B})$-separable density matrices~(\ref{sep}),
first observe that, thanks to
the Cauchy-Schwartz inequality, one has
$\sum_{k=0}^N\,p_k \langle J_{\vec{n}}\rangle^2_k\geq \langle J_{\vec{n}}\rangle^2$, where $\langle X\rangle^2_k$
denotes the mean-value of $X$ with respect to the number state $\vert k\rangle$.
Then,
\begin{eqnarray*}
\xi_S^2&\geq& N\,\frac{\sum_{k=0}^N p_k\, \Delta_k^2 J_{\vec{n}_1}}
{\langle J_{\vec{n}_2}\rangle^2+\langle J_{\vec{n}_3}\rangle^2}\\
&=&
\frac{1-n_{1z}^2}{n_{2z}^2+n_{3z}^2}\,\frac{\sum_{k=0}^Np_k\,N\Big(N+2k(N-k)\Big)}{\Big(
\sum_{k=0}^Np_k\,(N-2k)\Big)^2}\\
&\geq&\frac{\sum_{k=0}^N p_k\left(k-\frac{N}{2}\right)^2}
{\left(\sum_{k=0}^N p_k\left(k-\frac{N}{2}\right)\right)^2}\ ,
\end{eqnarray*}
where the last inequality follow from the second line of~(\ref{xiS1}).
A further application of the Cauchy-Schwartz inequality to the right hand side of the last inequality
yields $\xi^2_S\geq 1$ for all $(\mathcal{A},\mathcal{B})$-separable states when $n_1\neq z$.

If one chooses $\vec{n}_1=\hat{z}$, in the case of $(\mathcal{A},\mathcal{B})$-separable mixed states,
one finds $\Delta^2 J_z\neq 0$ and $\langle J_{\vec{n}_2,\vec{n}_3}\rangle=0$; therefore,
$\xi_S^2$ (and $\xi^2_W$) diverges.
Instead, for $(\mathcal{A},\mathcal{B})$-separable pure states $\vert k\rangle$
also $\Delta_kJ_z=0$ whence $\xi_S^2$ (and $\xi^2_W$) are not defined and must thus be computed
by means of suitable limiting procedures.

Let us consider the $(\mathcal{A},\mathcal{B})$-entangled vector state
$\vert \Psi\rangle=\sum_{k=0}^N \sqrt{p_k}\,\vert k\rangle$, with
real coefficients from a probability distribution $\pi=\{p_k\}_{k=0}^N$ over the stochastic variable $k$.
Then, from~(\ref{FI1})--(\ref{FI3}) it follows that $\langle J_y\rangle=0$ and $\Delta^2J_z=\Delta^2k$;
therefore $\xi_S^2=\xi^2_W$ and
\begin{equation}
\label{aux}
\xi^2_W=\frac{N\,\Delta^2J_z}{\langle J_x\rangle^2}=
\frac{N\,\Delta^2 k}{\Big(\sum_{k=1}^N\sqrt{k(N-k+1)}\,\sqrt{p_k\,p_{k-1}}
\Big)^2}\ .
\end{equation}
In the case of a Gaussian distribution peaked around $k=\ell\neq 0,N$,
\begin{equation}
\label{xigauss0}
p_k=\frac{1}{Z}\,\exp\left(-\frac{(k-\ell)^2}{\sigma^2}\right)\ ,\
Z=\sum_{k=0}^N\exp\left(-\frac{(k-\ell)^2}{\sigma^2}\right)\ ,
\end{equation}
one finds
\begin{equation}
\label{xigauss1}
\xi_W^2=\frac{2N+\mathcal{O}\left(e^{-\frac{1}{2\sigma^2}}\right)}
{\left(\sqrt{(\ell+1)(N-\ell)}+\sqrt{\ell(N-\ell+1)}\right)^2}\ .
\end{equation}
Thus, for sufficiently small $\sigma$, $\xi_W^2<1$ for all $\ell\neq 0,N$.

On the other hand, by choosing
\begin{equation}
\vert\Psi\rangle=\frac{p}{N}\sum_{k=0\,,\,k\neq N/2}^N\vert k\rangle+(1-p)\,\vert N/2\rangle\ ,\quad
0<p<1\ ,
\label{xigeq1_ent}
\end{equation}
it turns out that $\Delta^2k=p(N+2)(N+1)/12$ so that~(\ref{aux}) yields
$$
\xi^2_W=\frac{N(N+1)}{12(
\sqrt{1-p}+q)^2}\ ,
$$
where
$$
q=\sqrt{\frac{p}{N^2(N+2)}}\sum_{k\neq N/2,N/2+1}\sqrt{k(N-k+1)}\ .
$$
Letting $p\to0$ one gets $\vert\psi\rangle\to\vert N/2\rangle$ and,  if $N>3$,
$$
\xi^2_W\to N(N+1)/12>1\ .
$$

\noindent
\textbf{Remark 3.}\quad
The above two examples show that, when $\vec{n}_1=\hat{z}$, the spin-squeezing parameters
$\xi_{W,S}$ are not well-defined: different values for $\xi^2_{W,S}$ can be obtained by approaching
a state $\vert k\rangle$ via different limit procedures.
This fact is also of practical importance: indeed,
in~\cite{Oberthaler2}, approximations to Fock states $\vert k\rangle$ have been experimentally constructed
that are characterized by spin-squeezing parameters $\xi^2_W<1$.
This property arises from the fact that the approximations
are $(\mathcal{A},\mathcal{B})$-entangled states.
The previous discussion shows that some care has to be taken in constructing the perturbations
of $\vert k\rangle$; indeed, not all $(\mathcal{A},\mathcal{B})$-entangled states arbitrarily
close to it automatically have $\xi^2_{W,S}\geq 1$.
Therefore, the spin-squeezing parameters $\xi^2_{W,S}$ are not always useful for metrological applications,
a better quantity is the so-called quantum Fisher information~\cite{Braunstein1996}, which as we shall show below,
is continuous and well defined for all Bosonic qubits.

\subsection{Quantum Fisher Information}

In a measurement of the angle $\theta$ based on the state rotation~(\ref{state-rot}),
the error $\Delta\theta$  given by a locally unbiased estimator $E$ of the angle $\theta$ is bounded
by (see the Appendix)
\begin{equation}
\label{Fisher1}
\Delta^2\theta\geq \frac{1}{F[\rho,J_{\vec{n}_1}]}\ ,
\end{equation}
where $F[\rho,J_{\vec{n}_1}]$ is the so-called quantum Fisher information associated with the
rotation of $\rho$ around $\vec{n}_1$.

In order to overcome the shot-noise limit $\Delta^2\theta=1/N$, the quantum
Fisher information must then be strictly
larger than $N$.
In the appendix it is also showed that, in full generality,
\begin{equation}
\label{Fisher2}
F[\rho,J_{\vec{n}_1}]\,\Delta^2 J_{\vec{n}_2}\geq\langle J_{\vec{n}_3}\rangle^2\ ,
\end{equation}
where $\vec{n}_{1,2,3}$ is a triplet of orthogonal spatial directions.
Thus, if $\langle J_{\vec{n}_3}\rangle\neq 0$, one gets the following relation between the quantum Fisher
information and the squeezing parameter $\xi^2_W$ in~(\ref{Squeezparam}):
\begin{equation}
\label{Fisher3}
F^{-1}[\rho,J_{\vec{n}_1}]\leq \frac{\Delta^2J_{\vec{n}_2}}{\langle J_{\vec{n}_3}\rangle^2}=\frac{\xi_W^2}{N}\ .
\end{equation}
In the case of distinguishable qubits, from~(\ref{Fisher1}) and~(\ref{Fisher3}) it follows that spin-squeezing,
namely $\xi^2_W<1$, opens the possibility of achieving
$\Delta^2\theta<1/N$, thus of beating the shot-noise limit.

In the case of identical qubits and of $(\mathcal{A},\mathcal{B})$-separable states,
the right hand side of the above inequality diverges if $\vec{n}_1=\hat{z}$ as $\langle J_{\vec{n}_3}\rangle=0$,
while it does not make sense if $\vec{n}_2=\hat{z}$ for then also $\Delta^2J_{\vec{n}_2}=0$
whence, as already observed, $\xi^2_W$ is not defined.
However, the quantum Fisher information is always well-defined.
Indeed, using~(\ref{A4}) and~(\ref{FI3}), one finds that, if $k\neq 0,N$,
\begin{equation}
\label{conc1}
F[\vert k\rangle\langle k\vert,J_{\vec{n}}]=4\,\Delta^2J_z=(1-n^2_z)\Big(N+2k(N-k)\Big)>N
\end{equation}
for
$\displaystyle0\leq n^2_z<\frac{2k(N-k)}{N+2k(N-k)}<1$.
In particular,
$$
F[\vert k\rangle\langle k\vert,J_y]=N+2k(N-k)\ ,
$$
so that, according to~(\ref{Fisher1}),
for all $k\neq 0,N$ the $(\mathcal{A},\mathcal{B})$-separable pure states $\vert k\rangle$
might overcome the shot-noise limit, with the twin Fock state $\vert N/2\rangle$ yielding
$F[\rho,J_y]=\mathcal{O}(N^2)$ thus permitting to approach the so-called Heisenberg limit
$\Delta^2\theta=1/N^2$.
\medskip

\noindent
\textbf{Remark 4.}\quad
Notice that even approximating a number state $\vert \ell\rangle$ by an experimentally
more realistic superposition
$\vert \ell,\sigma\rangle$ of states $\vert k\rangle$
with coefficients as in~(\ref{xigauss0}), may beat the shot noise limit.
Indeed, one computes
$$
F[|\ell,\sigma\rangle\langle \ell,\sigma\vert,J_y]=4\,\Delta^2J_y=N+2\ell(N-\ell)+
\mathcal{O}\left(e^{-\frac{1}{\sigma^2}}\right)\ ,
$$
which can be kept $>N$  by suitably small $\sigma$.

Instead, making the quantum Fisher information larger than $N$ is impossible without
$(\mathcal{A},\mathcal{B})$-non-locality; indeed, $F[\vert \ell\rangle\langle \ell\vert,J_z]=4\,\Delta^2 J_z=0$.
Even considering the $(\mathcal{A},\mathcal{B})$-entangled perturbation $\vert \ell,\sigma\rangle$
does not help; indeed,
$$
F[\vert \ell,\sigma\rangle\langle\ell,\sigma\vert,J_z]=4\,\Delta^2 J_z=8 {\rm e}^{-\frac{1}{\sigma^2}}+
\mathcal{O}\left({\rm e}^{-\frac{2}{\sigma^2}}\right)\ .
$$
Therefore, the lower bound to the error in~(\ref{Fisher1})
becomes arbitrarily large when $\sigma\to 0$.
\medskip

When dealing with $(\mathcal{A},\mathcal{B})$-separable mixed states~(\ref{sep}), by means
of equation~(\ref{A3}) in the Appendix, one computes~\cite{Benatti2010}
\begin{eqnarray}
\noindent
F[\rho, J_{\vec{n}}]&=&
(1-n^2_z)\Big(N+2N\langle k\rangle-\langle k^2\rangle\\
\label{Fishdens}
&-&
4\sum_{k=0}^N {p_k p_{k+1}\over p_k+p_{k+1}}
(k+1)(N-k)\Big)\ .
\end{eqnarray}
Thus, if~(\ref{conc1}) holds for a certain $\vert\ell\rangle$, then, by continuity,
$F[\rho, J_{\vec{n}}]>N$ for a probability distribution $\pi=\{p_k\}_{k=0}^N$ suitably peaked around $k=\ell$, hence
able to overcome the shot-noise limit.

On the other hand, from the previous section we know that for all such mixed states
$\xi_W^2\geq \xi^2_S\geq 1$; therefore, based on this lower bound to the squeezing parameter, we would wrongly discard
such states as not useful for metrological applications.

\section{Discussion}

Unlike for distinguishable qubits,
in the case of identical Bosonic qubits, entangled states are not necessary to reach sub-shot noise accuracies in
parameter estimation.
This phenomenon is surely due to the non-local character of the system; however, the non-locality is
not in the states $\vert k\rangle$ which in fact are $(\mathcal{A},\mathcal{B})$-separable, rather in the
$(\mathcal{A},\mathcal{B})$-non-local character of the state-rotation~(\ref{state-rot}) generated by
$J_{\vec{n}}\neq J_z$.
Were the qubits distinguishable, neither the state nor the rotation would carry elements of non-locality
so that in order to beat the shot-noise limit, the state should be turned into an entangled one
before feeding the interferometric apparatus practically implementing the state-rotation~(\ref{state-rot}).
This is exactly what is done via a spin-squeezing technique in the experiments reported in~\cite{Oberthaler1,Treutlein}.
Instead, the main point we make here is that in experiments involving identical qubits, no preliminary
squeezing is needed before rotating the state. One might as well do with, say, a state
$\vert N/2\rangle$, the rotation around $\vec{n}\neq\hat{z}$ taking care of introducing the necessary non-locality.

\section{Appendix}

A most used quantum Fisher information $F[\rho,J_{\vec{n}}]$ is given by~\cite{Luo}
\begin{equation}
\label{A1}
F[\rho,J_{\vec{n}}]:={\rm tr}\Big(\rho\,L^2\Big)\ ,
\end{equation}
where $L$, known as symmetric logarithmic derivative, is a Hermitean operator such that
\begin{equation}
\label{A2}
\partial_\theta\rho_\theta\Big\vert_{\theta=0}=\frac{\rho\,L\,+\,L\,\rho}{2}=-i\,[J_{\vec{n}}\,,\,\rho]\ .
\end{equation}
Given a spectral decomposition $\rho=\sum_j r_j\,\vert r_j\rangle\langle r_j\vert$, one computes
\begin{equation}
\label{A3}
F[\rho,J_{\vec{n}}]=2\,\sum_{i,j\,:\,r_i\neq r_j}\frac{(r_i-r_j)^2}{r_i+r_j}\,
\Big|\langle r_i\vert J_{\vec{n}}\,\vert r_j\rangle\Big|^2\ .
\end{equation}
From such an expression one sees that the quantum Fisher information is a  continuous function
of the state $\rho$ and that, for pure states,
\begin{equation}
\label{A4}
F[\vert\psi\rangle\langle\psi\vert,J_{\vec{n}}]=4\,\Delta^2_\psi J_{\vec{n}}\ .
\end{equation}
An estimator $E$ is locally unbiased if
$\displaystyle\partial_\theta{\rm Tr}(\rho_\theta\,E)\Big\vert_{\theta=0}=1$;
then, inequality~(\ref{Fisher1}) follows from applying to this relation the Cauchy-Schwartz
inequality for matrices
$$
\Big|{\rm Tr}(AB)\Big|^2\leq {\rm Tr}(A^\dag\,A){\rm Tr}(B^\dag\,B)\ .
$$
Analogously, inequality~(\ref{Fisher2}) follows from the fact that
${\rm Tr}(\rho_\theta(J_{\vec{n}_2}-\langle J_{\vec{n}_2}\rangle_\theta)=0$ implies
$$
\Big\vert{\rm Tr}\Big(\partial_\theta\rho_\theta(J_{\vec{n}_2}-\langle J_{\vec{n}_2}\rangle_\theta\Big)\Big\vert=
\Big\vert\partial_\theta\langle J_{\vec{n}_2}\rangle_\theta\Big\vert=
\Big\vert\langle J_{\vec{n}_3}\rangle_\theta\Big\vert\ .
$$

\end{document}